\documentclass[aps,prl,amsmath,showpacs,amssymb,twocolumn]{revtex4}
\usepackage{amssymb}
\usepackage{color}
\usepackage{graphicx}
\usepackage{subfigure}
\usepackage{epsfig,amssymb,amsmath}
\usepackage[hypertex,linkcolor=red]{hyperref}

\usepackage{color}

\def\>{\rangle}
\def\<{\langle}
\def\comment#1{}
\def\labell#1{\label{#1}}
\def\section#1{{\par\em #1:--- }}
\def\togli#1{}

\begin{document}

\title{Stronger Uncertainty Relations for the Sum of Variances}
\author{Lorenzo Maccone$^1$ and Arun K. Pati$^{2,3}$} \affiliation{
  \vbox{$^1$Dip.~Fisica and INFN Sez.~Pavia, University ~of Pavia, via
    Bassi 6, I-27100 Pavia, Italy}\\\vbox{$^2$Quantum Information and
    Computation Group, Harish-Chandra Research Institute, Chhatnag
    Road, Jhunsi, Allahabad 211 019, India\\$^3$Department of Mathematics,
Zhejiang University, Hangzhou 310027, PR~China}}

\begin{abstract}
  Heisenberg-Robertson's uncertainty relation expresses a limitation
  in the possible preparations of the system by giving a lower bound to
  the product of the variances of two observables in terms of their
  commutator.  Notably, it does not capture the concept of
  incompatible observables because it can be trivial, i.e., the lower
  bound can be null even for two non-compatible observables. Here we
  give two stronger uncertainty relations, relating to the sum of
  variances, whose lower bound is guaranteed to be nontrivial whenever
  the two observables are incompatible on the state of the system.
  \comment{Note that incompatible observables can have a common
    eigenvector, so this last specification is necessary}
\end{abstract}
\maketitle

In his seminal paper \cite{heis,zur} Heisenberg analyzes various
notions of uncertainties for measurement of non-commuting observables
in quantum theory. Here we deal with Robertson's formalization
\cite{robertson} that implies a restriction on the possible
preparations of the properties of a system.  Indeed, the
Heisenberg-Robertson uncertainty relation quantitatively expresses the
impossibility of jointly sharp preparation of incompatible
observables.  However, in practice, the conventional uncertainty
relations cannot achieve this, because the lower bound in the
uncertainty relation inequalities can be null and hence trivial even
for observables that are incompatible on the state of the system
(namely, the state is not a common eigenstate of both observables).
This is due to the fact that the uncertainty relations are expressed
in terms of the product $\Delta A^2\Delta B^2$ of the variances of the
measurement results of the observables $A$ and $B$, and the product
can be null even when one of the two variances is different from zero.
Here we provide a different uncertainty relation, based on the sum
$\Delta A^2 +\Delta B^2$, that is guaranteed to be nontrivial whenever
the observables are incompatible on the state.

Uncertainty relations are useful for a wide range of applications that
span from the foundations of physics all the way to technological
applications: they are useful for formulating quantum mechanics
\cite{lathibusch} (e.g.~to justify the complex structure of the
Hilbert space \cite{lahti} or as a fundamental building block for
quantum mechanics and quantum gravity \cite{hall}), for entanglement
detection \cite{guhne,olga}, for the security analysis of quantum key
distribution in quantum cryptography (e.g.~see \cite{peresfuchs}),
etc.  Previous uncertainty relations that provide a bound to the sum
of the variances comprise a lower bound in terms of the variance of
the sum of observables \cite{arun}, a lower bound based on the
entropic uncertainty relations \cite{huang}, and a sum uncertainty
relation for angular momentum observables \cite{angel}. In contrast to
the last, our bound applies to general observables, and in contrast to
the previous ones, it is built to be strictly positive if the
observables are incompatible on the state of the system.

\section{Stronger uncertainty relations} The Heisenberg-Robertson uncertainty
relation \cite{robertson} bounds the product of the variances through
the expectation value of the commutator
\begin{eqnarray}
\Delta A^2\Delta B^2\geqslant\left|\tfrac 12\<[A,B]\>\right|^2
\labell{hrun}\;,
\end{eqnarray}
where the expectation value and the variances are calculated on the
state of the quantum system $|\psi\>$. It was strengthened by
Schr\"odinger \cite{schroedinger} who pointed out that one can add an
anti-commutator term, obtaining
\begin{eqnarray}
\Delta A^2\Delta B^2\geqslant\left|\tfrac 12\<[A,B]\>\right|^2+
\left|\tfrac 12\<\{A,B\}_+\>-\<A\>\<B\>\right|^2
\labell{schun}\;.
\end{eqnarray}
Both these inequalities can be trivial even in the case in which $A$
and $B$ are incompatible on the state of the system $|\psi\>$, e.g.~if
$|\psi\>$ is an eigenstate of $A$, all terms in \eqref{hrun} and
\eqref{schun} vanish. Both relations can be derived through an
application of the Cauchy-Schwarz inequality.

A simple lower bound for the sum of the variances can be obtained from
these, by noticing that $(\Delta A-\Delta B)^2\geqslant 0$, so that,
using \eqref{hrun}, we find $\Delta A^2+\Delta B^2\geqslant 2\Delta
A\Delta B\geqslant |\<[A,B]\>|$. This inequality is still not useful,
as the lower bound can be null even if $A$ and $B$ are
incompatible on $|\psi\>$ so that the sum is trivially bounded as
$\Delta A^2+\Delta B^2>0$.  Instead, the following two inequalities
(which are the main result of this paper) have lower bounds which are
nontrivial. The first inequality is
\begin{eqnarray}
\Delta A^2+\Delta B^2\geqslant\pm i\<[A,B]\>+\left|\<\psi|A\pm
  iB|\psi^\perp\>\right|^2
\labell{lorun}\;,
\end{eqnarray}
which is valid for arbitrary states $|\psi^\perp\>$ orthogonal to the state of
the system $|\psi\>$, where the sign should be chosen so that $\pm
i\<[A,B]\>$ (a real quantity) is positive.  The lower bound in
\eqref{lorun} is nonzero for almost any choice of $|\psi^\perp\>$ if
$|\psi\>$ is not a common eigenstate of $A$ and $B$
(Fig.~\ref{f:fig}): just choose $|\psi^\perp\>$ that is orthogonal to
$|\psi\>$ but not orthogonal to the state $(A\pm iB)|\psi\>$. Such a
choice is always possible unless $|\psi\>$ is a joint eigenstate of
$A$ and $B$.

For illustration, we give an example of how one can choose
$|\psi^\perp\>$: if $|\psi\>$ is an eigenstate of $A$ one can choose
$|\psi^\perp\>=(B-\<B\>)|\psi\>/\Delta B\equiv|\psi^\perp_B\>$ (see
below), or $|\psi^\perp\>=(A-\<A\>)|\psi\>/\Delta
A\equiv|\psi^\perp_A\>$ if $|\psi\>$ is an eigenstate of $B$. If
$|\psi\>$ is not an eigenstate of either and
$|\psi^\perp_A\>\neq|\psi^\perp_B\>$, one can choose
$|\psi^\perp\>\propto(\openone-|\psi^\perp_B\>\<\psi^\perp_B|)|\psi^\perp_A\>$,
or $|\psi^\perp\>=|\psi^\perp_A\>$ if
$|\psi^\perp_A\>=|\psi^\perp_B\>$.
An optimization of $|\psi^\perp\>$ (namely, the choice that maximizes
the lower bound), will saturate the inequality \eqref{lorun}: it
becomes an equality.

A second inequality with nontrivial bound even if $|\psi\>$ is
an eigenstate either of $A$ or of $B$ is
\begin{eqnarray}
\Delta A^2+\Delta B^2\geqslant\tfrac12|\<\psi_{A+B}^\perp|A+B|\psi\>|^2
\labell{arun}\;,
\end{eqnarray}
where $|\psi^\perp_{A+B}\>\propto(A+B-\<A+B\>)|\psi\>$ is a state
orthogonal to $|\psi\>$ (with $\<O\>$ denoting the expectation value
of $O$). The form of $|\psi^\perp_{A+B}\>$ implies that the
right-hand-side of \eqref{arun} is nonzero unless $|\psi\>$ is an
eigenstate of $A+B$.

Clearly, both inequalities \eqref{lorun} and \eqref{arun} can be
combined in a single uncertainty relation for the sum of variances:
\begin{eqnarray}
\Delta A^2+\Delta B^2\geqslant\max({\cal L}_{(3)},{\cal L}_{(4)})
\labell{unc}\;,
\end{eqnarray}
with ${\cal L}_{(3),(4)}$ the right-hand-side of \eqref{lorun} and
\eqref{arun}, respectively.

Some comments on \eqref{lorun} and \eqref{arun} follow: (i)~they
involve the sum of variances, so one must introduce some dimensional
constants in the case in which $A$ and $B$ are measured with different
units; (ii)~removing the last term in \eqref{lorun}, we find the
inequality $\Delta A^2+\Delta B^2\geqslant |\<[A,B]\>|$ implied by the
Heisenberg-Robertson relation, as shown above; (iii)~using the same
techniques employed to derive \eqref{lorun}, one can also obtain an
amended Heisenberg-Robertson inequality:
\begin{eqnarray}
\Delta A\Delta B\geqslant\pm \tfrac i2\<[A,B]\>\Big/\Big(1-\frac
12\Big|\<\psi|\frac A{\Delta A}\pm i\frac B{\Delta B}|\psi^\perp\>\Big|^2\Big)
\labell{hrunc2}\;,
\end{eqnarray}
which reduces to \eqref{hrun} when minimizing the lower bound over
$|\psi^\perp\>$ and becomes an equality when maximizing it.

\begin{figure}[hbt]
\begin{center}
\epsfxsize=.7\hsize\leavevmode\epsffile{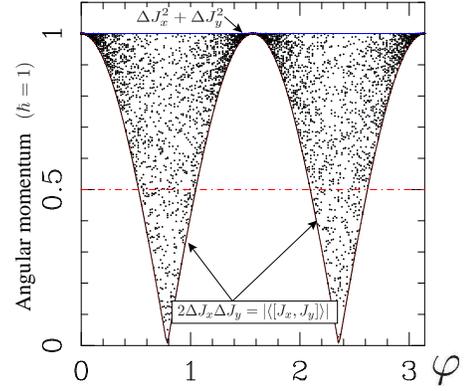}
\end{center}
\vspace{-.5cm}
\caption{Example of comparison between the Heisenberg-Robertson
  uncertainty relation \eqref{hrun} and the new ones \eqref{lorun},
  \eqref{arun}.  We choose $A=J_x$ and $B=J_y$, two components of the
  angular momentum for a spin 1 particle, and a family of states
  parametrized by $\varphi$ as
  $|\psi\>=\cos\varphi|+\>+\sin\varphi|-\>$, with $|\pm\>$ eigenstates
  of $J_z$ corresponding to the eigenvalues $\pm1$. None of these is a
  joint eigenstate of $J_x$ and $J_y$, nonetheless the
  Heisenberg-Robertson can be trivial for $\varphi=\pi/4$ and
  $\varphi=3\pi/4$. The lower curves are the product of the
  uncertainties and the expectation value of the commutator (this is a
  favorable case for the Heisenberg-Robertson relation since the
  product of uncertainties and its lower bound coincide). The upper
  curve is $\Delta J_x^2+\Delta J_y^2=1$ (it is constant for this
  family of states). The dash-dotted line is the bound \eqref{arun},
  the black points are the calculation of the bound \eqref{lorun} for
  20 randomly chosen states $|\psi^\perp\>$ for each of the 200 values
  of the phase $\varphi$ depicted. It is clear that the bound
  \eqref{lorun} well outperforms the Heisenberg-Robertson one for
  almost all choices of $|\psi^\perp\>$.  [The random $|\psi^\perp\>$
  are generated by generating a random unitary $U$ (uniform in the
  Haar measure) using the procedure detailed in \cite{zyc}, applying
  it to the $|+\>$ state, projecting on the orthogonal subspace to
  $|\psi\>$, and renormalizing the resulting state. Namely
  $|\psi^\perp\>\propto(\openone-|\psi\>\<\psi|)U|+\>$.]}
\labell{f:fig}\end{figure}

\section{Proofs of the results} In this section we provide two proofs
of the proposed uncertainty relations \eqref{lorun}, \eqref{arun}, and
\eqref{hrunc2}. The first proof, based on the parallelogram law, was
communicated to us by an anonymous Referee, while the second
(independent) proof was our original argument. While the first proof
is preferable because of its simplicity, we retain also the second for
completeness.

To prove \eqref{lorun}, define $C\equiv A-\<A\>$, $D\equiv B-\<B\>$ so
$\Delta A=\|C|\psi\>\|$, $\Delta B=\|i D|\psi\>\|$, where the
imaginary unit $i$ is introduced for later convenience. We have
\begin{eqnarray}
\|(C\mp iD)|\psi\>\|^2=\Delta A^2+\Delta B^2\mp i\<[A,B]\>
\labell{p3}\;, 
\end{eqnarray}
where the left-hand-side can be lower bounded through the Schwarz
inequality as
\begin{eqnarray}
&&|\<\psi|(A\pm iB)|\psi^\perp\>|^2=
|\<\psi|A\pm iB-\<A\pm iB\>|\psi^\perp\>|^2\nonumber\\&&=
|\<\psi|C\pm iD|\psi^\perp\>|^2\leqslant\|(C\mp iD)|\psi\>\|^2
\labell{p4}\;,
\end{eqnarray}
valid for all $|\psi^\perp\>$ orthogonal to $|\psi\>$, whence
\eqref{lorun} follows by joining \eqref{p3} and \eqref{p4}. The
equality condition for \eqref{lorun} follows from the equality
condition of the Schwarz inequality, namely iff $|\psi^\perp\>\propto
(A\mp iB-\<A\mp iB\>)|\psi\>$.

Up to now we have considered only a pure state $|\psi\>$ of the
system. This relation can be extended to the case of mixed states
$\rho=\sum_jp_j|\psi_j\>\<\psi_j|$ at least in the case in which it is
possible to choose a $|\psi^\perp\>$ that is orthogonal to all states
$|\psi_j\>$ (in the other cases, it is still possible to use the
inequality, but it cannot be expressed as an expectation value for the
density matrix). For each state $|\psi_j\>$ we can write \eqref{lorun}
as
\begin{eqnarray}
&&\Delta A_j^2+\Delta B_j^2\geqslant\mp
i\mbox{Tr}([A,B]|\psi_j\>\<\psi_j|)\nonumber
\\&&+\mbox{Tr}[(-A\pm
iB)|\psi^\perp\>\<\psi^\perp|(-A\mp iB)|\psi_j\>\<\psi_j|]
\labell{d8}\;,
\end{eqnarray}
where $\Delta A_j^2$ and $\Delta B_j^2$ are the variances calculated
on $|\psi_j\>$. By multiplying both members by $p_j$ and summing over
$j$, we obtain the mixed-state extension of \eqref{lorun}:
\begin{eqnarray}
&&\Delta A^2+\Delta B^2\geqslant\mp
i\<[A,B]\>\nonumber\\&&+\<(-A\pm
iB)|\psi^\perp\>\<\psi^\perp|(-A\mp iB)\>
\labell{d9}\;.
\end{eqnarray}

To prove \eqref{arun} we use the parallelogram law in Hilbert space to obtain
\begin{eqnarray}
  2\Delta A^2+2\Delta B^2=\|C+ \alpha D|\psi\>\|^2+
  \|C- \alpha D|\psi\>\|^2,\label{parall}
\end{eqnarray} 
for $C=A-\<A\>$, $D=B-\<B\>$, and $\alpha \in\mathbb{C}$ with
$|\alpha|=1$.  Since $\Delta(A+B)=\|(C+D)|\psi\>\|$,
$\Delta(A-B)=\|(C-D)|\psi\>\|$, Eq.~\eqref{parall} for $\alpha=1$ is
equal to
\begin{eqnarray}
  \Delta A^2+\Delta  B^2&=&
\tfrac 12[\Delta(A+B)^2+\Delta(A-B)^2]\nonumber\\&\geqslant&
\tfrac 12\Delta(A+B)^2,
\labell{p2}\;
\end{eqnarray}
which is equivalent to \eqref{arun} since
$\Delta(A+B)^2=|\<\psi^\perp_{A+B}|A+B|\psi\>|^2$. The equality
condition for \eqref{arun} is immediate from \eqref{p2}: $|\psi\>$
must be an eigenstate of $A-B$.  Also, note that the lower bound in
\eqref{arun} is nonzero unless $|\psi\>$ is an eigenstate of $A+B$.
Clearly $|\psi\>$ can be an eigenstate of $A+B$ without being an
eigenstate of either $A$ or $B$, but in the interesting case when
$|\psi\>$ is an eigenstate of one of the two (which trivializes both
Heisenberg's and Schr\"odinger's uncertainty relations), the lower
bound must be nonzero unless $|\psi\>$ is an eigenstate of both.
It is also easy to use \eqref{p2} to modify the inequality
\eqref{arun} so that it has always a nontrivial lower bound except
when $|\psi\>$ is a joint eigenstate of $A$ and $B$, namely
\begin{eqnarray}
\Delta A^2+\Delta
  B^2&\geqslant&\max(\tfrac12|\<\psi_{A+B}^\perp|A+B|\psi\>|^2,
  |\<\psi_{A}^\perp|A|\psi\>|^2,\nonumber\\
&&|\<\psi_{B}^\perp|B|\psi\>|^2)
\labell{a3}\;.
\end{eqnarray}
[Note that one can also obtain \eqref{lorun} from the parallelogram law
\eqref{parall} for $\alpha = \pm i$.]

We now provide a second proof of \eqref{lorun} and \eqref{arun}, and a
proof of \eqref{hrunc2}. They use the square-modulus inequality and
follow a procedure analogous to the one employed by Holevo to derive
the following useful relation \cite{holevo}:
\begin{eqnarray}
  \Delta A+\Delta A' \geqslant (a-a')|\<\psi|\psi'\>|\Big/
  \sqrt{2(1-|\<\psi|\psi'\>|)}
\labell{holunc}\;,
\end{eqnarray}
where $a$, $a'$ are the expectation values of $A$ on the states
$|\psi\>$ and $|\psi'\>$ respectively, $\Delta A^2$ and $\Delta A'^2$
are the variances on the same states.

To derive \eqref{lorun} start from the inequality
\begin{eqnarray}
&&\!\!\!\!\!\!\!\!\!
\|c_A\epsilon(A-a)|\psi\>\pm
ic_B(B-b')|\psi'\>+c(\epsilon|\psi\>-|\psi'\>)\|^2\geqslant 0,
\nonumber
\\&&\labell{d1}
\end{eqnarray}
with $a=\<\psi|A|\psi\>$, $b'=\<\psi'|B|\psi'\>$,
$\epsilon\equiv\<\psi|\psi'\>/|\<\psi|\psi'\>|$, and $c_A$, $c_B$, and
$c$ real constants. Calculating the square modulus, we find
\begin{eqnarray}
c_A^2\Delta A^2+c_B^2\Delta B'^2\geqslant-c^2\gamma-c_Ac_Bc\delta\mp
ic_Ac_B\kappa
\labell{d2}\;,
\end{eqnarray}
with $\Delta A^2$ and $\Delta B'^2$ the variances of $A$ and $B$ on
$|\psi\>$ and $|\psi'\>$ respectively, and where
$\gamma\equiv2(1-|\<\psi|\psi'\>|)$, $\delta\equiv
2$Re$(\epsilon^*\<\psi|a-A\pm i(B-b')|\psi'\>)$, and
$\kappa\equiv2i$Im$(\epsilon^*\<\psi|(A-a)(B-b')|\psi'\>)$. Now choose
the value of $c$ that maximizes the right-hand-side of \eqref{d2}
(assuming that one chooses the sign so the last term is positive),
namely $c=-c_Ac_B\delta/(2\gamma)$. Whence, inequality \eqref{d2}
becomes
\begin{eqnarray}
  c_A^2\Delta A^2+c_B^2\Delta B'^2\geqslant (c_Ac_B\delta)^2/(4\gamma)\mp
  ic_Ac_B\kappa
\labell{d3}\;.
\end{eqnarray}
Depending on the choice of $c_A$ and $c_B$ one can prove \eqref{lorun}
or \eqref{hrunc2}. Start with the former by taking $c_A=c_B=1$, we
find
\begin{eqnarray}
&&  \Delta A^2+\Delta B'^2\geqslant \frac{\delta^2}{4\gamma}\mp
  i\kappa=
\frac{[\mbox{Re}(\epsilon\<\psi'|(-\bar A\mp i\bar
  B')|\psi\>)]^2}{2(1-|\<\psi|\psi'\>|)}
\nonumber\\&&
\mp i(\epsilon^*\<\psi|\bar A\bar
  B'|\psi'\>-\epsilon\<\psi'|\bar B'\bar A|\psi\>)
\;,\
\labell{d4}\;
\end{eqnarray}
where $\bar A\equiv A-a$ and $\bar B'\equiv B-b'$. This inequality,
which may be of independent interest, is a two-observable extension of
the Holevo inequality \eqref{holunc}, and reduces to it by choosing
$\bar B=\pm i(A-a')$\comment{CHECK THIS!} and recalling that $(\Delta
A+\Delta A')^2\geqslant\Delta A^2+\Delta A'^2$. To obtain
\eqref{lorun}, take the limit $|\psi'\>\to|\psi\>$. This can be
calculated by writing
$|\psi'\>=\cos\alpha|\psi\>+e^{i\lambda}\sin\alpha|\psi^\perp\>$,
where $|\psi^\perp\>$ is orthogonal to $|\psi\>$ and taking the limit
$\alpha\to0$. The arbitrariness of $|\psi'\>$ ensures the
arbitrariness of $|\psi^\perp\>$ and of the phase $\lambda$. In the
limit, the last term of \eqref{d4} yields the expectation value of the
commutator and the other term on the right-hand-side tends to
$[$Re$(e^{i\lambda}\<\psi|(-A\pm i B)|\psi^\perp\>)]^2$. For either
signs in this expression, we can choose $\lambda$ so that the term in
parenthesis is real, so that this expression can be written also as
$|\<\psi|(-A\pm i B)|\psi^\perp\>|^2$. This implies that the limit
$|\psi'\>\to|\psi\>$ of \eqref{d4} gives \eqref{lorun} (with the above
choice of $\lambda$).

To prove the second proposed uncertainty relation \eqref{hrunc2}, we
can choose $c_A=\Delta B'$ and $c_B=-\Delta A$ in \eqref{d3}, which
then becomes
\begin{eqnarray}
&&  \Delta A\Delta B'\geqslant\pm\tfrac i2(\epsilon^*\<\psi|\bar A\bar
  B'|\psi'\>-\epsilon\<\psi'|\bar B'A|\psi\>)\nonumber\\&&+
\frac{\Delta A\Delta
    B'}{4(1-|\<\psi|\psi'\>|)}\Big[\mbox{Re}\Big(\epsilon^*\<\psi|\frac{\bar
    A}{\Delta A}\pm i\frac{\bar B'}{\Delta B'}|\psi'\>\Big)\Big]^2
\labell{d6}\!\!\!\!.
\end{eqnarray}
We can now take the limit $|\psi'\>\to|\psi\>$ using the same
procedure described above. Again the first term tends to the
expectation value of the commutator, while the second term tends to
${\Delta A\Delta B}[$Re$(e^{-i\lambda}\<\psi^\perp|A/\Delta A\mp
iB/\Delta B|\psi\>)]^2/2$. Again the phase $\lambda$ can be chosen so
that this last term is real and \eqref{d6} becomes
\begin{eqnarray}
 \Delta A\Delta B\geqslant\pm\tfrac i2\<[A,B]\>+\tfrac{\Delta A\Delta
   B}2\Big|\<\psi^\perp|\tfrac A{\Delta A}\mp i\tfrac B{\Delta B}|\psi\>\Big|^2
\nonumber\labell{d7}\;,
\end{eqnarray}
which is equivalent to \eqref{hrunc2}.

Finally, the second proof of \eqref{arun} is obtained by noting that
$(\Delta A+\Delta B)^2 \le 2 (\Delta A^2+\Delta B^2)$. Therefore, we
have
\begin{eqnarray}
&&\Delta A^2+\Delta B^2 \ge  \tfrac12[\Delta (A+B)]^2
\labell{a1}\;,
\end{eqnarray}
where we have used the sum uncertainty relation of \cite{arun}, namely
$\Delta A+\Delta B\geqslant\Delta(A+B)$ with $[\Delta(A+B)]^2$ the
variance of $(A+B)$ in the state $ |\psi\>$. The meaning of the sum
uncertainty relation is that mixing different operators always
decreases the uncertainty.  The lower bound in \eqref{a1} can be
rewritten using Vaidman's formula \cite{vaidman}
\begin{eqnarray}
O|\psi\>=\<O\>|\psi\>+\Delta O|\psi^\perp_O\>
\labell{vaid}\;,
\end{eqnarray}
(the expectation value $\<O\>$ and the variance $\Delta O^2$ of
the observable $O$ are calculated on $|\psi\>$), obtaining
\begin{eqnarray}
\Delta O=|\<\psi^\perp_O|\Delta O|\psi^\perp_O\>|=
|\<\psi^\perp_O|(O-\<O\>)|\psi\>|=
|\<\psi^\perp_O|O|\psi\>|,\nonumber
\end{eqnarray}
which, inserted into \eqref{a1} with $O= (A+ B)$ gives \eqref{arun}. 
Using the results of \cite{arun} it is also easy to extend this
inequality to more than two observables.

\section{Possible choices of $|\psi^\perp\>$}
We now show that the optimization over $|\psi^\perp\>$ of both
inequalities \eqref{lorun} and \eqref{hrunc2} makes them tight. Start
with \eqref{lorun}: the lower bound is clearly maximized if we choose
$|\psi^\perp\>$ as close as possible to the state $|\chi\>=(A\pm
iB)|\psi\>$, for example projecting such state into the orthogonal
subspace to $|\psi\>$ as
$|\psi^\perp\>=(\openone-|\psi\>\<\psi|)|\chi\>/{\cal N}$, with $\cal
N$ a normalization. With this choice, we find \begin{eqnarray}
  && \<\psi^\perp|(A\pm iB)|\psi\>=\<\psi|[A-a\mp
  i(B-b)]\times\labell{d10}\\&&\nonumber
(A\pm iB)|\psi\>/{\cal N}=(\Delta A^2+\Delta B^2\pm i\<[A,B]\>)/{\cal N}
\;,
\end{eqnarray}
where the normalization constant is ${\cal N}=(\Delta A^2+\Delta
B^2\pm i\<[A,B]\>)^{1/2}$. Substituting \eqref{d10} into
\eqref{lorun}, we see that the inequality is indeed saturated.
Analogous considerations hold for \eqref{hrunc2}: in this case, we
should choose $|\psi^\perp\>\propto(\openone-|\psi\>\<\psi|)(\tfrac
A{\Delta A}\mp i\tfrac B{\Delta B}|\psi\>$. With this choice,
$\<\psi^\perp|(\tfrac A{\Delta A}\mp i\tfrac B{\Delta B}|\psi\>=2\mp
i\<[A,B]\>/(\Delta A\Delta B)$, which is also equal to the square of
the normalization constant for $|\psi^\perp\>$. Hence, substituting
this value in \eqref{hrunc2}, we see that it is saturated for this
choice of $|\psi^\perp\>$. [It is also clear that the choice of
$|\psi^\perp\>$ that minimizes the lower bounds transforms
\eqref{lorun} into $\Delta A^2+\Delta B^2\geqslant |\<[A,B]\>|$ that
is a consequence of \eqref{hrun} as shown above, and it transforms
\eqref{hrunc2} into \eqref{hrun}.]

A simple prescription for how to choose an expression for
$|\psi^\perp\>$ uses \eqref{vaid}, namely
$|\psi^\perp\>=(O-\<O\>)|\psi\>/\Delta O$.

Here we have focused on extending the Heisenberg-Robertson uncertainty
relation \eqref{hrun}, but it is also possible to give an extension to
the Schr\"odinger relation \eqref{schun}, by choosing an arbitrary
phase factor $e^{i\theta}$ in place of the imaginary constant $i$ in
\eqref{d1}.

\section{Uncertainty relations and uncertainty principle}
Recently, there has been an interesting and lively debate on how to
interpret the uncertainty principle \cite{ozawa,werner}. To elucidate
the relation between these results and ours, we introduce Peres'
nomenclature that distinguishes between uncertainty {\em relation} and
uncertainty {\em principle} \cite{peresbook}.  The former refers
solely to the preparation of the system which induces a spread in the
measurement outcomes, and does not refer to the disturbance induced by
the measurement or to joint measurements \footnote{A good definition
  of ``uncertainty {\em relation}'' is given in \cite{peresbook}, pg.
  93: ``The only correct interpretation of [the uncertainty relations
  for $x$ and $p$] is the following: If the same preparation procedure
  is repeated many times, and is followed either by a measurement of
  $x$, or by a measurement of $p$, the various results obtained for
  $x$ and for $p$ have standard deviations, $\Delta x$ and $\Delta p$,
  whose product cannot be less than $\hbar / 2$. There never is any
  question here that a measurement of $x$ 'disturbs' the value of $p$
  and vice-versa, as sometimes claimed.  These measurements are indeed
  incompatible, but they are performed on {\it different} particles
  (all of which were identically prepared) and therefore these
  measurements cannot disturb each other in any way.  The uncertainty
  relation [...] only reflects the intrinsic randomness of the
  outcomes of quantum tests.'' We emphasize that the uncertainty {\em
    relation} must not be confused with the uncertainty {\em
    principle}.}. The latter entails also the measurement disturbance
by the apparatus and the impossibility of joint measurements of
incompatible observables. From Robertson's derivation
\cite{robertson}, it is clear \cite{peresbook} that the
Heisenberg-Robertson inequalities are uncertainty {\em relations} (the
ones typically taught in textbooks). In contrast, Heisenberg in his
paper \cite{heis,zur} does not give a clear distinction between the
two concepts, and both can be applied depending on the systems he
analyzes there. The recent literature \cite{ozawa,werner} discusses
the uncertainty {\em principle}: measurement-induced disturbance and
joint measurability. Our result instead refers to uncertainty {\em
  relations}: it can be seen as a quantitative expression for the
nonexistence of common eigenstates in incompatible observables.

\togli{We also comment on the relationship to complementarity. The most
radical departure from the classical world is embodied by Bohr's
principle of complementarity \cite{bohr}, but its obscure and
non-quantitative formulation hinders its fruition: often uncertainty
relations are  preferred.  Complementarity and uncertainty are
different concepts: complementarity (loosely) refers to the
impossibility of precisely determining incompatible observables,
whereas uncertainty {\em in principle} refers to the impossibility of
precisely determining incompatible observables while measuring a
system in a given state.  However, {\em in practice}, conventional
formulations of uncertainty do not express this, because the
uncertainty relations can be trivial, namely, the lower bound in the
inequality can be null even for observables that are incompatible on
the state of the system. The uncertainty relations provided here
bridge part of the gap between uncertainty and complementarity.}

\section{Conclusions}
The Heisenberg-Robertson \eqref{hrun} or Schr{\"o}\-din\-ger
\eqref{schun} uncertainty relations do not fully capture the
incompatibility of observables on the system state. In this paper, we
have presented a stronger uncertainty relation \eqref{unc} based on
two lower bounds \eqref{lorun} and \eqref{arun} for the sum of the
variances that are nontrivial if the two observables are incompatible
on the state of the system.  We also derived \eqref{hrunc2}, a
strengthening of the Heisenberg-Robertson uncertainty relation
\eqref{hrun}. \togli{These new additions to the quantum mechanics
  toolkit will have implications in foundational aspects as well as
  technological spin offs.}  There exists alternate formulations of
uncertainty relations in terms of bounds on the sum of {\em entropic}
quantities \cite{deutsch,mass}, but our new relations capture the
notion of incompatibilty in terms of experimentally measured error
bars, as they refer to variances.

\vskip1\baselineskip LM acknowledges useful discussions with
A.S.~Holevo and V.~Giovannetti. AKP thanks the project K.P.~Chair
Professor of Zhejiang University of China. We acknowledge the
contribution of an anonymous Referee that has provided the proof based
on the parallelogram law.


\begin{references}
\bibitem{heis} W. Heisenberg, 
  Zeit. Phys. {\bf 43}, 172 (1927), English translation in \cite{zur},
  pg.  62--84.
\bibitem{zur} J. A. Wheeler, H. Zurek, {\em Quantum Theory and
    Measurement}, (Princeton Univ.~Press, Princeton, 1983).
\bibitem{robertson}H. P. Robertson, 
  Phys.  Rev. {\bf 34}, 163 (1929).
\bibitem{lathibusch}P. Busch, T. Heinonen, P. J. Lahti, 
  Physics Reports {\bf 452}, 155 (2007).
\bibitem{lahti}P. J. Lahti, M. J. Maczynski, J. Math. Phys. {\bf
    28}, 1764 (1987).
\bibitem{hall} M. J. W. Hall, 
  Gen. Rel. Grav. {\bf 37}, 1505 (2005).
\bibitem{guhne}O. G\"uhne, Phys. Rev. Lett. {\bf 92}, 117903 (2004).
\bibitem{olga}H. F. Hofmann, S. Takeuchi, 
  Phys. Rev. A {\bf 68}, 032103 (2003).
\bibitem{peresfuchs} C. A. Fuchs, A. Peres, 
  Phys. Rev. A {\bf 53}, 2038 (1996).
\bibitem{arun} A. K. Pati, P. K. Sahu, 
  Phys. Lett. A {\bf 367}, 177 (2007).
\bibitem{huang} Y. Huang, 
  Phys.  Rev. A {\bf 86}, 024101 (2012).
\bibitem{angel} A. Rivas, A. Luis, 
  Phys. Rev. A {\bf 77}, 022105 (2008).
\bibitem{schroedinger}E. Schr\"odinger, 
  Sitzungsberichte der Preussischen Akademie der Wissenschaften,
  Physikalisch-mathematische Klasse {\bf 14}, 296 (1930).
\bibitem{zyc} K.~\u Zyczkowski, P.~Horodecki, A.~Sanpera, M.~Lewenstein,
  Phys. Rev. A {\bf 58}, 883 (1998).
\bibitem{holevo} A. S. Holevo, 
  Teor.  Veroyatnost. i Primenen., {\bf 18}, 371 (1973), English
  translation in Theory Probab. Appl. {\bf 18}, 359 (1973).
\bibitem{vaidman} L.~Vaidman, 
  Am. J. Phys. {\bf 60}, 182 (1992).
\bibitem{ozawa} M. Ozawa, Phys. Rev. A {\bf 67}, 042105 (2003); M.
  Ozawa, Int. J. Quant. Inf. {\bf 1}, 569 (2003); M. Ozawa, Found.
  Phys. {\bf 41}, 592 (2011); M. Ozawa, AIP Conf. Proc. {\bf 1363}, 53
  (2011); J. Erhart et al., Nature Phys. {\bf 8}, 185 (2012); L. A.
  Rozema et al., Phys. Rev. Lett.  {\bf 109}, 100404, (2012); C. Branciard,
  Proc.  Natl. Acad. Sci. USA {\bf 110}, 6742 (2013); C. Branciard, Phys.
  Rev. A {\bf 89}, 022124 (2014).
\bibitem{werner}R.F. Werner, Quant.  Inform. Comput. {\bf 4}, 546
  (2004), quant-ph/0405184; L. Maccone, Europhys. Lett. {\bf 77},
  40002 (2007); P. Busch, P. Lahti, R.F. Werner, Phys.  Rev.  Lett.
  {\bf 111}, 160405 (2013); P. Busch, P. Lahti, R.F.  Werner, Phys.
  Rev. A {\bf 89}, 012129 (2014); P. Busch, P. Lahti, R.F.  Werner, J.
  Math.  Phys. {\bf 55}, 042111 (2014); F. Buscemi, M.J.W. Hall, M.
  Ozawa, M.M. Wilde, Phys. Rev. Lett. {\bf 112}, 050401 (2014).
\bibitem{peresbook}A. Peres, {\em Quantum Theory: Concepts and
    Methods}, (Kluwer ac. publ., Dordrecht, 1993).
\bibitem{deutsch} I. Bialynicki-Birula, J. Mycielski, Commun. Math.
  Phys. {\bf 44}, 129 (1975); D. Deutsch, Phys. Rev. Lett. {\bf 50},
  631 (1983).
\bibitem{mass}H. Maassen, J.B.M. Uffink, Phys. Rev. Lett. {\bf 60}, 1103
  (1988). 
\end{references}
\end{document}